# ABOUT CLASSICAL ASPECT OF THE CHIRAL SYMMETRY DYNAMICAL BREAKING IN THREE-DIMENSIONAL ELECTRODYNAMICS.

**M.SH.PEVZNER**

Department of Physics, National Mining University ofUkraine,
19, Karl Marx Avenu, , E-mail: PevznerM@nmuu.dp.ua, mark@omp.dp.ua
**DNIEPROPETROVSK, 49600, UKRAINE**

Abstract

Attention is paid to the fact that the stress tensor diagonal components of the point charged particle field in three-dimensional electrodynamics are equal to zero. It allows to suppose the particle mass origin is field in the model, if one loop correction to the classical potential of this field is taken into account. In the noticed approach static potential and field strength are evaluated and the electrical radius of the particle is found. It turns out that this radius is connected with the vacuum polarization and with the particle electrical field screening at the large distances.

-------

The dynamical symmetry breaking problem is of great importance in modern physics. It allows to considerate from one point of view the aggregation of the phenoma which at the first sight are not connected with each other. In quantum field theory the dynamical symmetry breaking leads to the appearance of particles masses if their "bare" masses are equal to zero [1-6]. This mechanism of mass appearance has an advantadge over the Higgs mechanism since it is not based on assumption of the existence of particles in nature which are not found till now. It stimulates the further investigation of this mechanism and of the field theory models where the particle mass arises as a result of dynamical symmetry breaking.

Three dimensional quantum electrodynamics ($QED_{2+1}$) is one of the models in which the dynamical chiral symmetry breaking takes place [7-11]. Interest to it may be explained by such circumstances. Firstly it is another field theory model in which at some approximation the dynamical chiral symmetry breaking takes place. Secondly there are some hopes that by through the use of this theory it will be possible to advance further in the understanding of some macroscopical effects, in particular of the high temperature superconductivity [12-14]. Thirdly at some approximation in $QED_{2+1}$ confinement takes place.

The problem considered here has the classical aspect too [15], but all attempts andertaken in this direction were unsuccessful. It seems the classical approach to the problem is meaningless since here the classical result is not the limiting case of the quantum one. However the results of the works [16,17] not to make this assertion so categorical. In other words it is advisable to return to the classical consideration of the self energy particle problem using to the quantum ideas only if the purely classical approach comes across some insuperable difficulties.

2. The main reason of the attempted failures of classical approach to the field mass particle problem consists in the fact that the stress tensor diagonal components of the particle field are not equal to zero. So the momentum components and the energy of this field do not form the Lorentz's vector what does not allow to consider this mass as having a field origin.

Let us notice that in the case of two spatial dimensions it is possible to avoid this problem at the classical level taking into account the vacuum polarization inclusion to the field potential formed by a particle. Then the admission about the field origin of the particle mass in given model will not call the objection.

Let us the frame system moves with the speed $v$ relative to the particle. Then for the time and spatial components of the field momentum we have [15]

$$P_1 = v/k \int \left(T_{44}^{(0)} - T_{11}^{(0)}\right)\left(d^2\vec{r}^{(0)}\right), \tag{1}$$

$$P_4 = i/k \int \left(T_{44}^{(0)} - v^2 T_{11}^{(0)}\right)\left(d^2\vec{r}^{(0)}\right). \tag{1a}$$

Here the index $^{(0)}$ signifies that the tensor density and elementary two-dimensional volume are taken in the particle rest system and besides the system of units with $\hbar = c\ 1, k = \sqrt{1-v^2}$ is used further.

Passing from the rest system to the moving one the particle momentum and energy have to transform according to low

$$P_1 = v/k \int T_{44}^{(0)}\left(d^2\vec{r}^{(0)}\right), \tag{2}$$

$$P_4 = i/k \int T_{44}^{(0)}\left(d^2\vec{r}^{(0)}\right). \tag{2a}$$

Thus as it follows from (1) - (2a) in order that momentum components and energy to form the Lorentz's vector, the condition

$$\int T_{11}^{(0)}\left(d^2\vec{r}^{(0)}\right) = 0 \tag{3}$$

has to be satisfied.

The stress tensor density is defined by the equality

$$T_{11}^{(0)} = E_1^{(0)^2} - (1/2)E^{(0)^2}. \tag{4}$$

By using the polar coordinate system we obtain

$$\int T_{11}^{(0)}\left(d^2\vec{r}^{(0)}\right) = \frac{1}{2}\int_{r_0}^{\infty} E^{(0)^2} r dr \int_0^{2\pi} \cos 2\varphi\, d\varphi, \tag{5}$$

here $r_0$ is the cutoff parameter which has the meaning of electrical radius. of a particle.

We shall notice that the presentation of the double integral as a repeated one a in (5) takes place only if the integral with respect to $r$ is convergent. In the classical case it does not take place in general but taking into account the vacuum polarization the situation may be changed. Then the integral with respect to the angle is equal to zero that leads to turning into zero of the left part of the equality (5).

3. We shall demonstrate the integral convergence with respect to the variable $r$ by direct calculations. With this purpose we shall evaluate field potential of a point charge and for that we shall proceed from the formula [18]

$$A_0(\vec{r}) = iQ/(2\pi)^2 \int D_R^{(t)}(\vec{k},0)e^{i\vec{k}\cdot\vec{r}}\left(d^2\vec{k}\right), \tag{6}$$

where $Q$ is a particle charge on the plane, $D_R^{(t)}(\vec{k},0)$ is the regularized transverse part of the full photon propagator at zero frequency. For this function we have

$$D_R^{(t)}(k) = 1/(ik^2 - \Pi_R(k^2));\qquad(7)$$

here $\Pi_R(k^2) = (1/(d-1))(\Pi_{\mu\mu}(k) - \Pi_{\mu\mu}(0))$, $\Pi_{\mu\nu}(k)$ is the polarization operator, $d$ is the space-time dimension. In our case $d = 3$.

In order to find $\Pi_R(k^2)$ it is necessary to solve the Shwinger-Dyson equations system for the propagators and vertexes in QED$_{2+1}$, but in common case it is a very complicated problem. So for its solving we shall restrict ourselves to the approximation which was used by studying the dynamical chiral breaking problem QED$_{2+1}$ [8-10]. It consists in using the $N^{-1}$ - approximation and in neglecting by the mass function by evaluating the polarization operator is. It gives [7]

$$\Pi_R(k^2) = -(i\alpha/8)(k^2)^{1/2},\qquad(8)$$

where $\alpha = e_0^2 N$, $e_0^2$ is the dimensional coupling constant at QED$_{2+1}$, $N$ is the fermions number. Further supposing that the chiral symmetry is broken dynamically at the $N = N_c$, we consider the mass function to be small enough near the critical meaning of the fermions number $N_c$. It allows us to act in the bifurcation theory spirit [19] and to consider relationship (8) as the first nonvanishing term of the polarization operator expansion into the functional series upon the mass function taking $N = N_c$ herein. So it is used the approximation which comes to the taking into account of only the fermion loop at the photon propagator in which fermion mass is equal to zero, and vertexes and the wavefunction renormalization are considered to be free. For the critical meaning of the fermions number in Landau gauge has been obtained the value $N_c \approx 3$ [8-10]. There are reasons to consider that the improvement of the approximation leads to the same result for the fermions number which does not depend on the gauge [20].

Using the formulae (6) - (8) and will fulfilling the integration with respect to the angle we will obtain

$$A_0(r) = Q/2\pi \int_0^\infty J_0(kr)/(k + \alpha/8) dk,\qquad(9)$$

where $J_0(kr)$ is the Bessel's function of zero order. Integration with respect to $k$ gives [21]

$$A_0(r) = (Q/4)(\mathbf{H}_0(\alpha r/8) - N_0(\alpha r/8)),\qquad(10)$$

where $\mathbf{H}_0$ and $N_0$ are respectevely Struve's and Neyman's functions of zero order.

Static potential $A_0(r)$ (10) has the integral representation [22]

$$A_0(r) = Q/2\pi \int_0^\infty e^{(-\alpha r/8)x}(1 + x^2)^{(-1/2)} dx,\qquad(11)$$

which allows to investigate its behaviour both at small ($\alpha r/8 \ll 1$) and at large ($\alpha r/8 \gg 1$) distances. Besides it follows from (11) that potential the $A_0(r)$ is a monotonic deacreasing function of the variable $r$. At large and small distances we have

$$A_0(r) \approx \begin{cases} (Q/2\pi)\ln(16/\alpha r), \text{if} & \alpha r/8 \ll 1, \\ (Q/2\pi)(8/\alpha r^2), \text{if} & \alpha r/8 \gg 1. \end{cases} \quad (12)$$

We shall notice the existence of the space domain in which $A_0 \sim 1/r$ is of crusial importance for presence of the dynamical chiral symmetry breaking in the case under consideration [23]. For the field strengh we have

$$E_r = -\frac{\partial A_0}{\partial r} = -\alpha Q/16\pi + (\alpha Q/32)(\mathbf{H}_1(\alpha r/8) - N_1(\alpha r/8)), \quad (13)$$

where $E_r$ is the nonzero projection of $\vec{E}$, $\mathbf{H}_1$ and $N_1$ are Struve's and Neyman's functions of the first order respectively. It is convenient to use also the integral representation for $E_r$

$$E_r = \alpha Q/16\pi \int_0^\infty e^{(-\alpha r/8)x} x \cdot (1+x^2)^{-1/2} dx, \quad (14)$$

from which we obtain

$$E_r \approx \begin{cases} Q/2\pi r, & \text{if} \quad \alpha r/8 \ll 1, \\ 8Q/2\pi\alpha r^2, & \text{if} \quad \alpha r/8 \gg 1; \end{cases} \quad (15)$$

Thus we see that the vacuum polarization changes the strengh behaviour so that with the distance arising it decreases more rapidly than in the case when the vacuum polarization is not taken into account. It may be cosidered as a result of the peculiar screening of the particle field in virtue of the vacuum polarization.

5. Let us evaluate a particle's self energy in our case. We shall proceed out of the expression for the self energy formed by the particle, which is in the rest

$$W = \int T_{44}^{(0)}(d^2\vec{r}^{(0)}) = \frac{1}{2}\int E_r^2(d^2\vec{r}^{(0)}). \quad (16)$$

Inserting of the expressions (15) into (16) gives

$$W \approx Q^2/4\pi \int_{r_0}^{8/\alpha} \frac{dr}{r} + 16Q/\pi\alpha^2 \int_{8/\alpha}^{\infty} \frac{dr}{r^3}. \quad (17)$$

If one proceeds from that the main contribution to the field energy gives the first term in (17) we obtain

$$W = (Q^2/4\pi)\ln(8/\alpha r_0). \quad (18)$$

We shall emphasize once again that the given expression cannot be obtained only remaining in the framework of the classical theory. It ought to be considered as a result of two cutoff: the particle electrical radius $r_0$ carries out the infrared cutoff and $8/\alpha$ carries out the ultraviolet one. It is an additional ground of the integrals cutoff in the momentum space which was carried out in the works [8-10].

Expression for the particle field mass has the form

$$m_0 = W = (Q^2/4\pi)\ln(8/\alpha r_0), \quad (18a)$$

from where it is possible to find its electrical radius

$$r_0 = (8/\alpha)e^{-4\pi m_0/Q^2}. \quad (19)$$

(In order to avoid misunderstanding we note that the quantities $W$ and $m_0$ are related to the plane and therefore they have dimensions of the linear energy density and linear mass density respectively).

In conclusion we note that radiation corrections are able to change the explicit form of the stress tensor density (4). Thus in order to examine the problem more carefully it is necessary to proceed from the nonlinear Lagrangian for three - dimensional electrodynamics.. In particular it is known in $QED_{3+1}$ the corrections to the point charge potential behaviour at the large distances evaluated by using of the nonlinear Lagrangian differ from those evaluated by using the relations (6) - (7) for the corresponding case [18]. One may think that such a situation takes place in $QED_{2+1}$. Therefore from our point of view it would be interesting to consider the point charge potential behaviour at the large distances, proceeding from the nonlinear Lagrangian in $QED_{2+1}$.

The author is greatly indebted to V.P.Gusynin for his attention to the work.